\documentclass[final,times,twocolumn]{elsarticle}
\usepackage{graphicx,amssymb,amsmath}
\usepackage{multirow,dcolumn,bm,latexsym,soul}
\newcolumntype{K}[1]{>{\centering\arraybackslash}p{#1}}

\journal{Physics Letters B}
\begin{document}
\newcommand{\be}{\begin{equation}}
\newcommand{\ee}{\end{equation}}
\newcommand{\bq}{\begin{eqnarray}}
\newcommand{\eq}{\end{eqnarray}}

\begin{frontmatter}

\title{Varying fine-structure constant cosmography}
\author[inst1,inst2]{C. J. A. P. Martins\corref{cor1}}\ead{Carlos.Martins@astro.up.pt}
\author[inst1,inst3]{F. P. S. A. Ferreira}\ead{up201907302@fc.up.pt}
\author[inst1,inst3]{P. V. Marto}\ead{up201907888@fc.up.pt}
\address[inst1]{Centro de Astrof\'{\i}sica da Universidade do Porto, Rua das Estrelas, 4150-762 Porto, Portugal}
\address[inst2]{Instituto de Astrof\'{\i}sica e Ci\^encias do Espa\c co, CAUP, Rua das Estrelas, 4150-762 Porto, Portugal}
\address[inst3]{Faculdade de Ci\^encias, Universidade do Porto, Rua do Campo Alegre 687, 4169-007 Porto, Portugal}

\cortext[cor1]{Corresponding author}

\begin{abstract}
Cosmography is a phenomenological and relatively model-independent approach to cosmology, where physical quantities are expanded as a Taylor series in the cosmological redshift, or in related variables. Here we apply this methodology to constrain possible cosmological variations of the fine-structure constant, $\alpha$. Two peculiarities of this case are the existence of high-redshift data, and the fact that one term in the series is directly and tightly constraint by local laboratory tests with atomic clocks. We use this atomic clock data, together with direct model-independent high-resolution astrophysical spectroscopy measurements of $\alpha$ up to redshift $z\sim7$ and additional model-dependent constraints on $\alpha$ from the cosmic microwave background and big bang nucleosynthesis, to place stringent (parts per million level) constraints on the first two terms in the $\alpha$ cosmographic series.
\end{abstract}

\end{frontmatter}

\section{Introduction}\label{into}

Mapping the expansion history of the universe is among the most pressing tasks of observational cosmology, especially considering the evidence for its recent acceleration phase. Usually this includes specific assumptions on an underlying model (or class of models), leading to cosmological constraints that are model dependent. A conceptual alternative to this, not yet realized in practice but expected to be achieved by forthcoming astrophysical facilities, are redshift drift measurements, also known as the Sandage test \cite{Sandage,Liske,Esteves}.

A further alternative, known as cosmography \cite{Visser}, relies in expanding physical quantities as a Taylor series, either in the cosmological redshift or in related variables, around the present time---that is, zero redshift. This removes the need for the \textit{a priori} choice of a specific model other than, say, assuming that it is of the Friedmann-Lema\^{\i}tre-Robertson-Walker class. One immediate limitation of a Taylor series in the redshift $z$ is that the series will diverge for $z>1$, but this can be avoided by expanding in the rescaled redshift \cite{Cattoen}, defined as
\be\label{defy}
y\equiv\frac{z}{1+z}\,.
\ee
Other limitations include the need to choose where the series is truncated, degeneracies between the series coefficients, and a dependence on chosen priors for these coefficients \cite{Dunsby}. Nevertheless, cosmography provides useful constraints on important quantities such as the decceleration and jerk parameters.

Spacetime variations of nature's fundamental couplings, such as the fine-structure constant $\alpha$, are expected (indeed, unavoidable) in most physically motivated extensions of our canonical theories of cosmology and particle physics. A recent review of the theoretical and observational status of the field can be found in \cite{ROPP}. Constraints on $\alpha$ at a given redshift are usually expressed relative to the present-day laboratory value $\alpha_0$, specifically,
\be\label{defa}
\frac{\Delta\alpha}{\alpha}(z)\equiv\frac{\alpha(z)-\alpha_0}{\alpha_0}\,.
\ee
The cosmological dependence of $\alpha$ in these models is of course model-dependent, but broadly speaking we expect, at least in the simplest (best motivated and scalar-field based) models, that $\alpha$ is constant during the radiation-dominated era, drifts slowly in the matter-dominated era, and then stabilizes again once the universe starts accelerating. In particular, the first of these behaviours will emerge when there is a coupling to the trace, $\rho-3p$, which will vanish in the radiation era. Examples of classes of models with this behaviour are \cite{Barrow,Damour}. Nevertheless, models with different behaviours can of course be constructed. Moreover, tight constraints on $\Delta\alpha/\alpha(z)$ exist on a wide range of redshifts, from local laboratory measurements to constraints and the big bang nucleosynthesis (BBN) epoch, and most of them are model-independent. These reasons motivate an exploration of a cosmographic approach to the putative redshift dependence of $\alpha$, which we start in this work.

The plan of the rest of this letter is as follows. We start in Sect. \ref{cosmog} by introducing the varying alpha cosmography series, and discussing the physical interpretation of its coefficients. Sect. \ref{data} describes the available datasets that we use, and Sect. \ref{differences} discusses the differences between the archival and dedicated high-resolution spectroscopy measurements of $\alpha$, which we treat separately because they are statistically incompatible with each other. This data is then used to constrain the series coefficients, with Sect. \ref{param2} presents the case where the series is truncated at quadratic order, while Sect. \ref{param3} discusses the differences when the series is instead truncated at cubic order. Finally, Sect. \ref{concl} presents some conclusions and outlines possible future steps.

\section{Varying alpha cosmography}\label{cosmog}

The simplest cosmographic approach would be to expand the relative variation of the fine-structure $\alpha$, defined in Eq. (\ref{defa}) as a Taylor series in the cosmological redshift
\be
\frac{\Delta\alpha}{\alpha}(z)=\frac{1}{\alpha_0}\left(\frac{d\alpha}{dz}\right)_0z+\frac{1}{2\alpha_0}\left(\frac{d^2\alpha}{dz^2}\right)_0z^2+\ldots\,,
\ee
where as before the index $0$ denotes the present-day value. However such a series need not converge at redshifts $z>1$ \cite{Cattoen}, so instead we work with the rescaled redshift $y$, defined in Eq. (\ref{defy})
\be\label{seriesy}
\frac{\Delta\alpha}{\alpha}(y)=\frac{1}{\alpha_0}\left(\frac{d\alpha}{dy}\right)_0y+\frac{1}{2\alpha_0}\left(\frac{d^2\alpha}{dy^2}\right)_0y^2+\ldots\,;
\ee
we will briefly discuss the issue of where to truncate the series later. Note that the first term in the series is the linear one: by definition there is no constant term.

Interestingly, we can write the first coefficient as
\be\label{defa1}
a_1\equiv\frac{1}{\alpha_0}\left(\frac{d\alpha}{dy}\right)_0=\frac{1}{\alpha_0}\left(\frac{d\alpha}{dz}\right)_0=-\frac{1}{H_0}\left(\frac{\dot\alpha}{\alpha}\right)_0\,,
\ee
where $H_0$ is the Hubble constant and the dot denotes a time derivative. The final term in brackets is the current drift rate of $\alpha$, which can be measured in a direct and model-independent way in local laboratory experiments, which yield very stringent constraints \cite{ROPP}, which we further discuss in the next section. For convenience in our subsequent analysis, we have defined this first coefficient in the $y$ series as $a_1$, and similarly we can define the second coefficient
\be\label{defa2}
a_2\equiv\frac{1}{\alpha_0}\left(\frac{d^2\alpha}{dy^2}\right)_0=\frac{1}{\alpha_0}\left(\frac{d^2\alpha}{dz^2}\right)_0+ \frac{2}{\alpha_0}\left(\frac{d\alpha}{dz}\right)_0\,.
\ee

To illustrate the information encoded by these coefficients, we provide two simple but representative examples. The first corresponds to a one-parameter redshift dependence of $\alpha$
\be
\frac{\Delta\alpha}{\alpha}(z)=\zeta\ln{(1+z)}\,;
\ee
this is often a good approximation for the behaviour of $\alpha$ in string theory inspired models, at least for low redshifts \cite{Damour,Vacher}, with $\zeta$ denotes the coupling of the scalar field (known, in this context, as the dilaton) to the electromagnetic sector of the theory. In this case the series coefficients are simply
\be
a_1=a_2=\zeta\,.
\ee
As has already been mentioned, $a_1$ will be tightly constrained by local atomic clock measurements. Whether or not the value of $a_2$ is consistent with $a_1$ will then provide a simple consistent test for this specific class of models.

As a second example, consider a canonical (quintessence type) scalar field, which is responsible for both the dark energy that is accelerating the universe and the spacetime variation of $\alpha$. The latter is again due to the coupling, still denoted $\zeta$, of the scalar field to the electromagnetic sector, and for simplicity we further consider a flat universe. In this case the redshift dependence of alpha is \cite{Calabrese}
\be
\frac{\Delta\alpha}{\alpha}(z) =\zeta \int_0^{z}\sqrt{3\Omega_\phi(z')\left[1+w_\phi(z')\right]}\frac{{\rm d}z'}{1+z'}\,,
\ee
where
\be
\Omega_\phi (z) \equiv \frac{\rho_\phi(z)}{\rho_{\rm tot}(z)} \simeq \frac{\rho_\phi(z)}{\rho_\phi(z)+\rho_{\rm m}(z)} \,,
\ee
is the fraction of the dark energy density (with the radiation density being neglected in the last step, on the assumption that one is working at low redshifts) and $w_\phi$ is the dark energy equation of state. In this case we have
\be
a_1=\zeta\sqrt{3(1-\Omega_m)(1+w_0)}
\ee
\be
a_2=\zeta\frac{\sqrt{3(1-\Omega_m)}}{\sqrt{(1+w_0)}}\left[\frac{1}{2}w'_0+ (1+w_0)\left(1+\frac{3}{2}\Omega_mw_0\right)\right]\,,
\ee
where $\Omega_m$, $w_0$ and $w'_0$ are respectively the present-day values of the matter density (as a fraction of the critical density), of the dark energy equation of state parameter, and of its derivative with respect to redshift. As expected in this class of models, measurements of $\alpha$ constrain the dark energy sector \cite{ROPP}.

\section{Available data}\label{data}

Here we discuss the available measurements of the fine-structure constant, to be used in the following sections to constrain the series coefficients $a_i$.

Firstly, as already mentioned, laboratory tests comparing atomic clocks based on transitions with different sensitivities to $\alpha$ lead to a constraint on the current drift rate which is, up to a minus sign, precisely the first series coefficient $a_1$. The most recent and most stringent such constraint is \cite{Lange}
\be
\left(\frac{\dot\alpha}{\alpha}\right)_0=(1.0\pm1.1)\times10^{-18}\,\text{yr}^{-1}\,.
\ee
Alternatively, and for a Hubble constant $H_0=70 km/s/Mpc$, this can be written
\be
\frac{1}{H_0}\left(\frac{\dot\alpha}{\alpha}\right)_0=0.014\pm0.015\,\text{ppm}\,;
\ee
note that for convenience this and analogous subsequent values are given in parts per million units. This measurement is model-independent, other than the choice of a value of the Hubble parameter, but the latter choice is clearly subdominant with respect to other components of our study's error budget.

The main part of our dataset consists of high-resolution spectroscopy measurements of $\alpha$, done (mainly at optical wavelengths) in low-density absorption clouds along the line of sight of bright quasars, typically with wavelength resolution $R=\lambda/\Delta\lambda\sim10\, 000-150\, 000$ (with the precise value being different for different measurements and spectrographs). Here we use two different subsets which we treat separately throughout, since as it will be seen presently the two are somewhat discrepant. The first is the \textit{Archival} dataset of Webb \textit{et al.} \cite{Webb}. The second is a compilation of \textit{Dedicated} measurements, including those listed in Table 1 of \cite{ROPP} and more recent ones from the Subaru telescope \cite{Cooksey} and the X-SHOOTER \cite{Wilczynska}, HARPS \cite{Milakovic} and ESPRESSO \cite{Welsh,Murphy} spectrographs. The former subset includes measurements up to redshift $z\sim4.18$ while the latter includes measurements up to redshift $z\sim7.06$, and the two subsets have comparable constraining power \cite{Meritxell}. 

Finally, our analysis will also include two additional measurements at higher redshift. The cosmic microwave background (CMB) provides a constraint on $\alpha$ at an effective redshift $z_{\rm CMB}\sim1100$. A recent analysis \cite{Hart} finds
\be
\left(\frac{\Delta\alpha}{\alpha}\right)_{\rm CMB}=(-0.7\pm2.5)\times10^3\,\text{ppm}\,.
\ee
We note that this constraint is doubly model-dependent, both because it assumes a specific cosmological model and because it makes the simplistic assumption that only $\alpha$ varies (and has a constant value at this epoch that may differ from the local laboratory one) while all remaining parameters have their standard values. In practical term this point is somewhat moot because the constraint is extremely weak (with an uncertainty at the parts per thousand level, as opposed to parts per million) and therefore has no statistical weight in our analysis, but we nevertheless include it for the sake of completeness.

Last but not least, BBN also provides a parts per million constraint, at an effective redshift $z_{\rm BBN}\sim4\times10^8$ \cite{Deal}
\be
\left(\frac{\Delta\alpha}{\alpha}\right)_{\rm BBN}=2.1^{+2.7}_{-0.9}\,\text{ppm}\,.
\ee
This is also a model-dependent constraint, but in a much milder sense than the CMB constraint. Specifically, it is a constraint in a broad class of Grand Unified Theories where all the couplings (gauge and Yukawa couplings, particle masses, etc.) are self-consistently allowed to vary, and with relevant phenomenological parameters allowed to vary and marginalized. It is also a comparatively conservative constraint in the sense that it only comes for the Deuterium and Helium-4 abundances; including Lithium-7 in the analysis would lead to a slightly larger shift from the null value. A more detailed discussion of this model dependence can be found in \cite{Deal}.

\section{Archival versus Dedicated data}\label{differences}

\begin{table}
\begin{center}
\caption{Basic properties of the two sets of high-resolution spectroscopy measurements of $\alpha$ used in our cosmographic analysis. Constraints are given at the one sigma ($68.3\%$) confidence level, while upper limits are given at the two sigma ($95.4\%$) confidence level.}
\label{table1}
\begin{tabular}{| c | c | c |}
\hline
Parameter & Archival & Dedicated \\
\hline
Weighted mean $\Delta\alpha/\alpha$ (ppm) & $-2.16\pm0.85$ & $-0.23\pm0.56$ \\
Weighted mean redshift $z_{\rm eff}$ & $1.50$ & $1.29$ \\
\hline
Pure dipole A (ppm) & $9.4\pm2.2$ & $<2.9$ \\
Logarithmic dipole A (ppm) & $9.9\pm2.3$ & $<3.3$ \\
\hline
\end{tabular}
\end{center}
\end{table}

Table \ref{table1} compares some simple characteristics of these two datasets. For the Archival dataset these have been previously reported in \cite{Pinho,ROPP}, while for the the Dedicated results we update the results therein considering the more recent measurements mentioned above. In the first two rows we effectively assume that there is a unique astrophysical value of $\Delta\alpha/\alpha$, which we estimate by taking the weighted mean of all the values in each dataset. Similarly, we identify an effective redshift of the sample by taking the weighted mean of the redshifts of the measurements in the set. We see that while the effective redshifts are comparable (reflecting the prevalence of stringent measurements around $z\sim1$), the archival data has a preference for a negative variation at more than two standard deviations, while the dedicated value is consistent with the null result.

\begin{figure*}
\begin{center}
\includegraphics[width=\columnwidth,keepaspectratio]{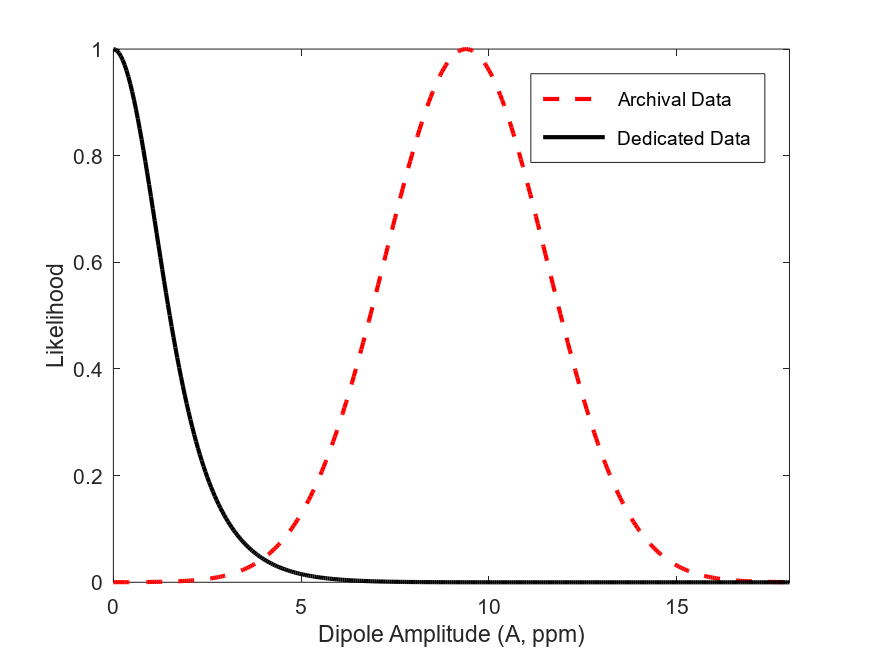}
\includegraphics[width=\columnwidth,keepaspectratio]{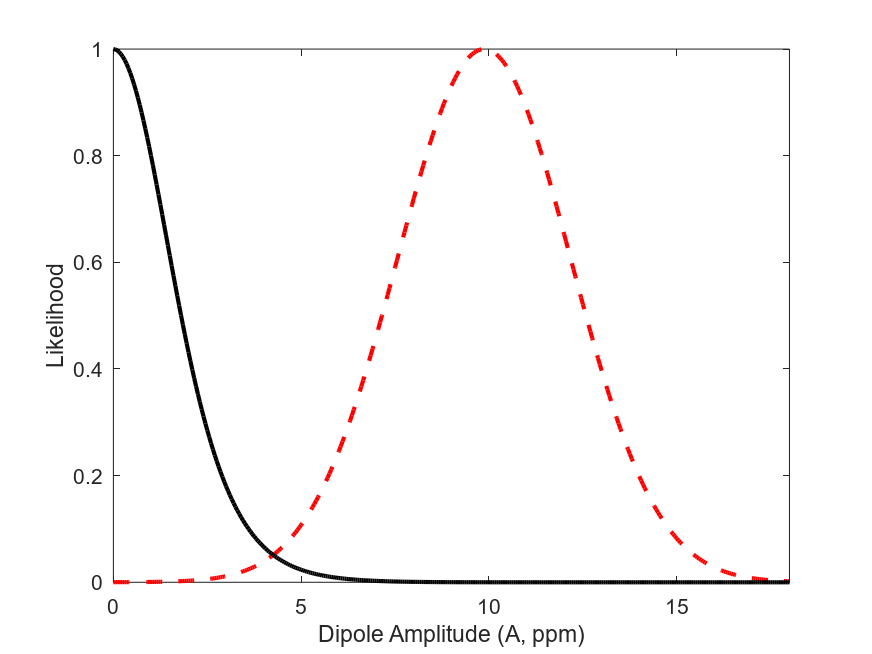}
\end{center}
\caption{\label{fig1}Posterior likelihood constraints on the amplitude (in parts per million), with the angular coordinates marginalized, of a putative dipole in the astrophysical measurements of $\alpha$. The left panel shows the result for a pure spatial dipole, while the right panel shows the result for a dipole with a further redshift dependence, cf. Eq. (\ref{puredipole}) and Eq. (\ref{redshiftdipole}) respectively. The red dashed and black solid lines correspond to the Archival and Dedicated spectroscopic datasets, described in the text.}
\end{figure*}

Moreover,  Webb \textit{et al.} \cite{Webb} find an indication of a spatial variation of $\alpha$ from the Archival dataset. It is straightforward to constrain the amplitude $A$ of a pure spatial dipole for the relative variation of $\alpha$
\be\label{puredipole}
\frac{\Delta\alpha}{\alpha}(A,\Psi)=A\cos{\Psi}\,,
\ee
which depends on the orthodromic distance $\Psi$ to the North Pole of the dipole (the locus of maximal positive variation) given by
\be\label{ortho}
\cos{\Psi}=\sin{\theta_i}\sin{\theta_0}+\cos{\theta_i}\cos{\theta_0}\cos{(\phi_i-\phi_0)}\,,
\ee
where $(\theta_i,\phi_i)$ are the Declination and Right Ascension of each measurement and $(\theta_0,\phi_0)$ those of the North Pole. The latter two coordinates, together with the overall amplitude $A$, are our free parameters. For comparison we further consider the case with a logarithmic redshift dependence in addition to the spatial variation
\be\label{redshiftdipole}
\frac{\Delta\alpha}{\alpha}(A,z,\Psi)=A\, \ln{(1+z)}\, \cos{\Psi}\,.
\ee
As previously mentioned such a logarithmic dependence is physically well motivated, and this parametrization has the further advantage of not requiring any additional free parameters. Constraints on the amplitude $A$, marginalizing over the angular coordinates, are shown in Table \ref{table1} and also Fig. \ref{fig1}. As is well know, for the Archival data there is a statistical preference, at just over four standard deviations, for a dipole with an amplitude of about nine parts per million, while in the Dedicated data there is no preference for a dipole, and the amplitude is constrained to be less than about three ppm at the $95.4\%$ confidence level. We note that the latter constraint improves the analogous ones reported in \cite{Pinho,ROPP} by about a factor of two, which highlights the impact of the growing number of dedicated measurements.

The above reasons therefore justify our separate treatment of the Archival and Dedicated datasets. The two separate analyses will also serve as an illustration of how the cosmographic parameters are affected by the two datasets.

\section{Two parameter cosmographic series}\label{param2}

We can now use the data described in the previous section to constrain the cosmographic series given in Eq. (\ref{seriesy}) for the rescaled redshift $y$. In other words, we assume that the series is truncated at the quadratic term. We separately consider the Archival and Dedicated high-resolution spectroscopy data (for the previously mentioned reasons). We further discuss separate results for what we call the \textit{Model-independent} dataset (which includes the atomic clocks and the spectroscopic data), and the \textit{Full dataset} (in which the CMB and BBN constraints are added to the former dataset).

\begin{figure*}
\begin{center}
\includegraphics[width=\columnwidth,keepaspectratio]{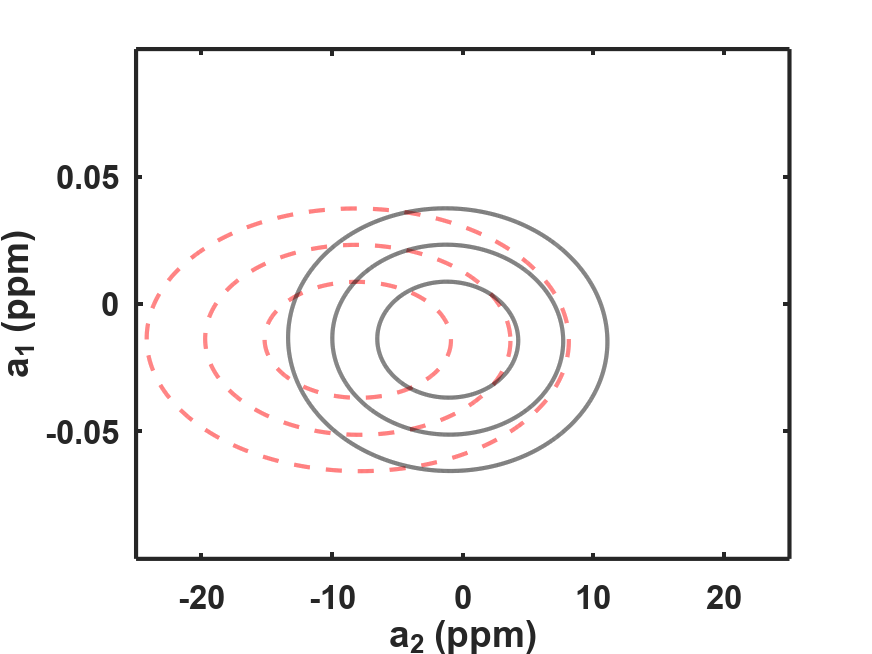}
\includegraphics[width=\columnwidth,keepaspectratio]{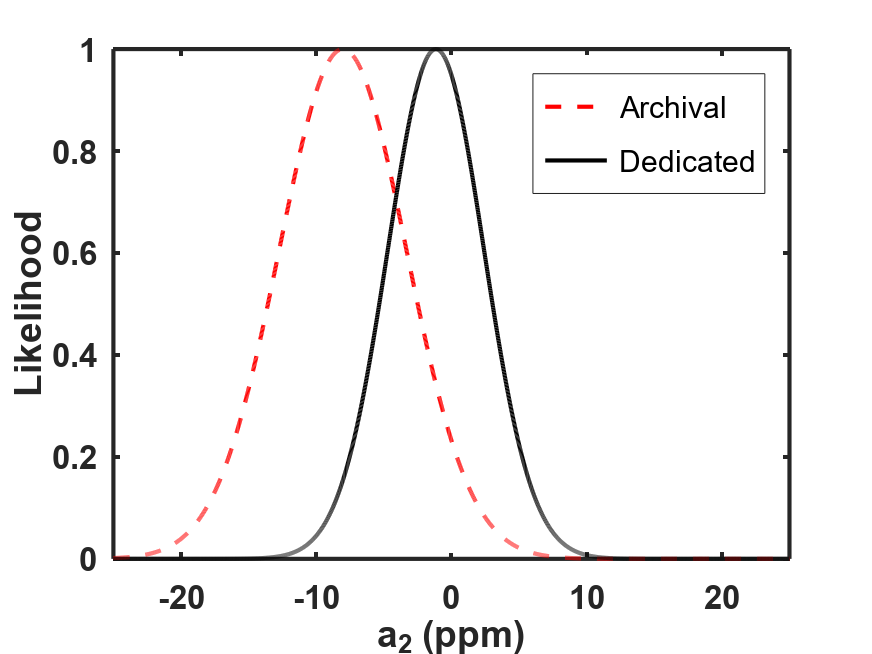}
\includegraphics[width=\columnwidth,keepaspectratio]{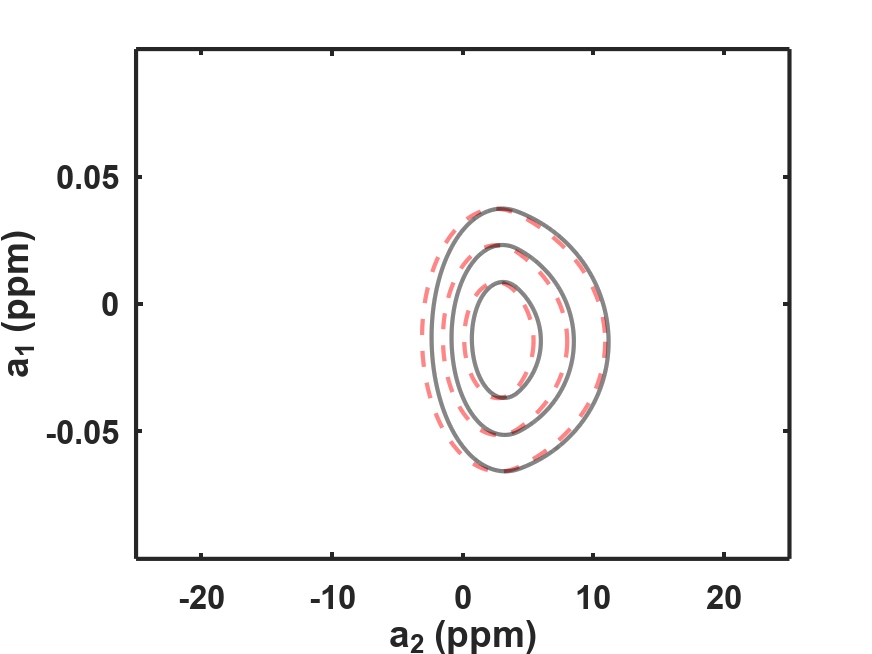}
\includegraphics[width=\columnwidth,keepaspectratio]{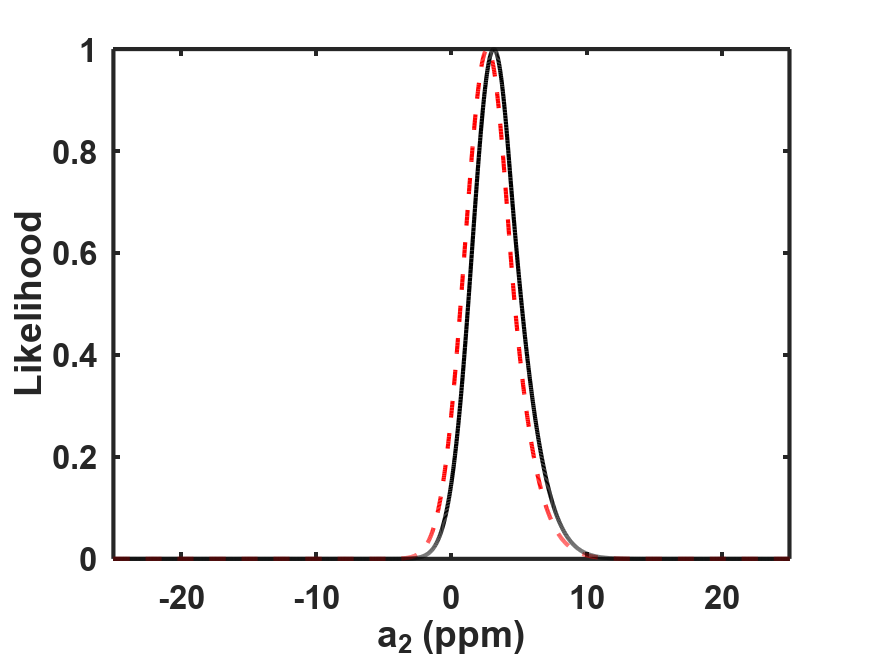}
\end{center}
\caption{\label{fig2}Constraints on the parameters $a_1$ and $a_2$, for a cosmographic series truncated at quadratic order. The left side panels show the one, two and three sigma confidence levels in the two-dimensional $a_1$--$a_2$ plane, while the right side panels show the posterior likelihood for $a_2$, with $a_1$ marginalized. The top and bottom row plots show the results for the model-independent and full datasets respectively. The red dashed and black solid lines correspond to the Archival and Dedicated spectroscopic datasets, described in the text, and parts per million units are used for all variables.}
\end{figure*}

The constraints obtained under these assumptions are depicted in Fig. \ref{fig2}. There are in principle two free parameters, $a_1$ and $a_2$, previously defined in Eqs. (\ref{defa1}--\ref{defa2}). However $a_1$ is directly constrained by the atomic clocks data, and since this is a very stringent constraint we find that the posterior likelihood for $a_1$ is always the same as the clocks constraint, except for the change of sign, \textit{i.e.}
\be
a_1=-0.014\pm0.015\,\text{ppm}\,;
\ee
therefore the one non-trivial parameter to be constrained by the data in this case is the coefficient of the quadratic term, $a_2$.

\begin{table}
\begin{center}
\caption{Posterior constraints on the cosmographic parameter $a_2$, with the remaining parameters marginalized. The first two rows are for the series truncated at quadratic order, while the last two rows are or the series truncated at cubic order. One sigma ($68.3\%$) confidence level constraints, in parts per million, are given throughout.}
\label{table2}
\begin{tabular}{| c | c | c | c |}
\hline
Series & Data & Archival & Dedicated \\
\hline
${\cal O}(y^2)$ & Model-independent & $-8.0\pm4.7$ & $-1.1\pm3.6$ \\
${\cal O}(y^2)$ & Full dataset & $+2.7\pm1.7$ & $+3.1\pm1.7$ \\
\hline
${\cal O}(y^3)$ & Model-independent & $-9.1_{-5.6}^{+5.8}$ & $-1.3_{-4.7}^{+4.8}$ \\
${\cal O}(y^3)$ & Full dataset & $-3.2_{-2.4}^{+3.9}$ & $-1.2_{-3.1}^{+4.7}$ \\
\hline
\end{tabular}
\end{center}
\end{table}

The first two rows of Table \ref{table2} list the posterior constraints on the $a_2$ parameter in the four relevant cases for the ${\cal O}(y^2)$ cosmographic series. For the archival data the statistical preference for a non-zero $a_2$ is less than two standard deviations, which further illustrates the impact of the atomic clocks constraint. On the other hand, for the dedicated data $a_2$ is consistent with zero, as expected. If one adds the CMB and BBN constraints, the latter's slight preference for a positive variation at high redshift pushes $a_2$ towards positive values for both the archival and dedicated spectroscopic datasets, but again the preference for a non-zero value is less than two standard deviations in both cases.

We thus see that when truncating the series at quadratic order the two coefficients are well constrained and seemingly without bias (in other words, values consistent with zero are recovered when expected). This behaviour is of course facilitated by the fact that the atomic clocks measurements directly constrains $a_1$. Moreover, the reduced chi-square is around $1.0$ for the archival dataset and around $0.8$ for the dedicated dataset, whether or not the high-redshift data  is used, so there is no obvious need for additional terms in the series. Nevertheless one may ask how the above results are affected when additional terms are included in the series. We address this point in the next section.

\section{Three parameter cosmographic series}\label{param3}

We now consider the case where the cosmographic series is extended to cubic order, in other words
\be
\frac{\Delta\alpha}{\alpha}(y)=\frac{1}{\alpha_0}\left(\frac{d\alpha}{dy}\right)_0y+\frac{1}{2\alpha_0}\left(\frac{d^2\alpha}{dy^2}\right)_0y^2+ +\frac{1}{6\alpha_0}\left(\frac{d^3\alpha}{dy^3}\right)_0y^3\,,
\ee
and we analogously define
\be
a_3\equiv\frac{1}{\alpha_0}\left(\frac{d^3\alpha}{dy^3}\right)_0\,.
\ee
We now have three free parameters. It is clear that $a_1$ will be well constrained as before, but we now expect a degeneracy between $a_2$ and $a_3$. The easy way to understand this is that in the limit $y\to1$ and assuming that there is no $\alpha$ variation we must approximately have $a_3=-3a_2$ (in writing this expression we have neglected $a_1$, which must be much smaller than ppm level). In other words, the BBN constrain on $\alpha$ will effectively constrain the combination $a_3+3a_2$, rather than the two parameters separately.

\begin{figure*}
\begin{center}
\includegraphics[width=\columnwidth,keepaspectratio]{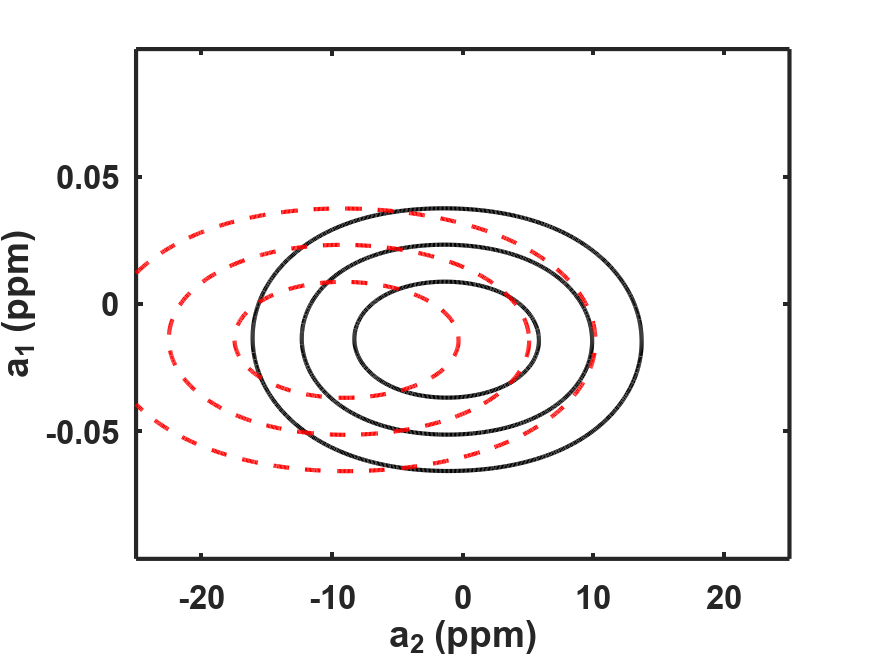}
\includegraphics[width=\columnwidth,keepaspectratio]{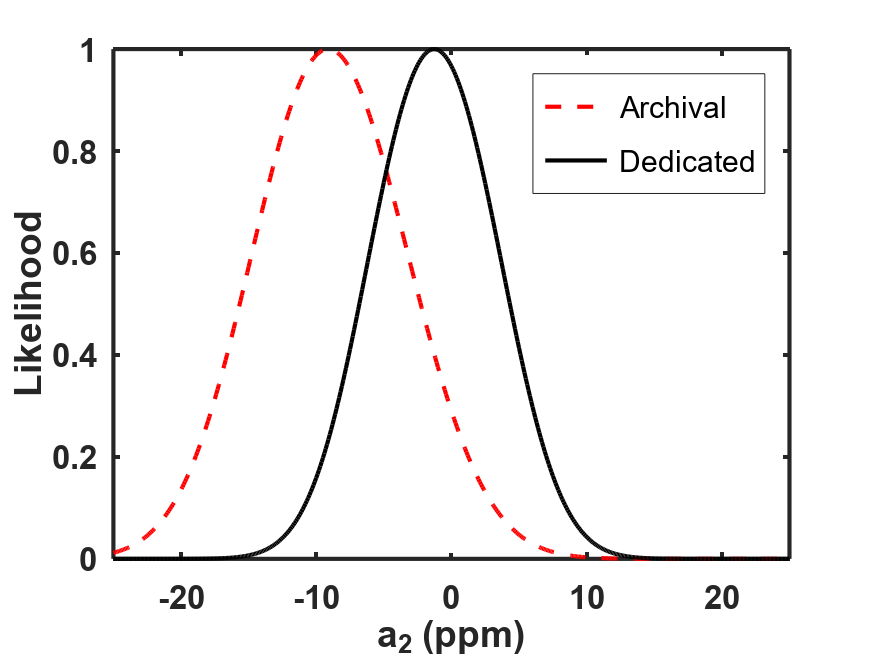}
\includegraphics[width=\columnwidth,keepaspectratio]{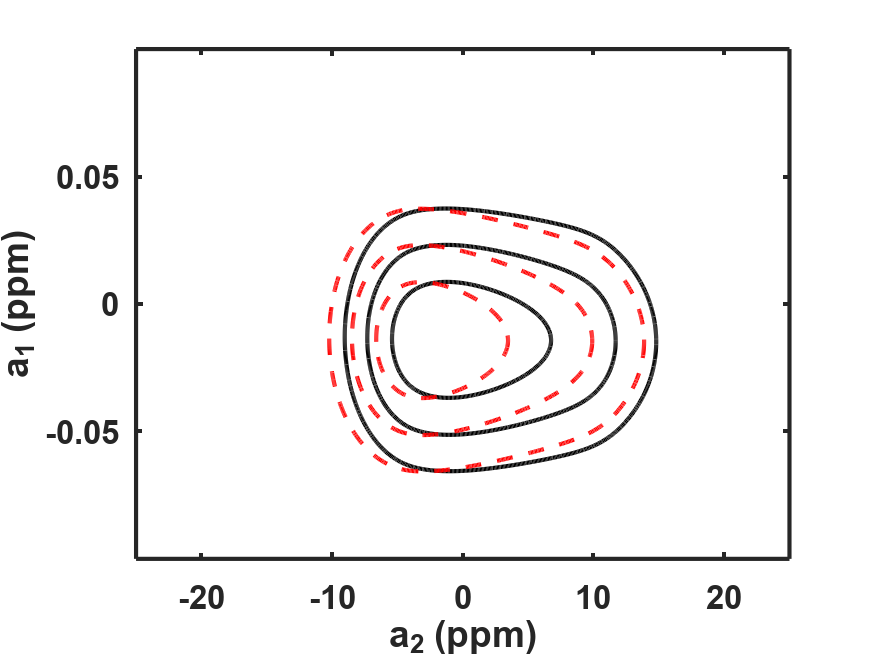}
\includegraphics[width=\columnwidth,keepaspectratio]{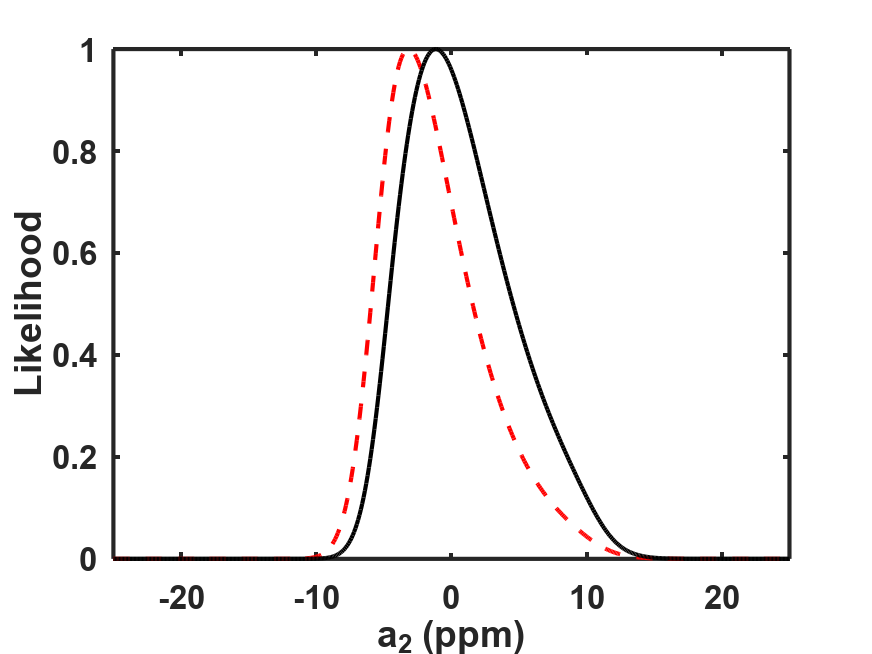}
\end{center}
\caption{\label{fig3}Constraints on the parameters $a_1$ and $a_2$, for a cosmographic series truncated at cubic order. The left side panels show the one, two and three sigma confidence levels in the two-dimensional $a_1$--$a_2$ plane, while the right side panels show the posterior likelihood for $a_2$, with $a_1$ marginalized; the cubic term coefficient, $a_3$, has been marginalized throughout. The top and bottom row plots show the results for the model-independent and full datasets respectively. The red dashed and black solid lines correspond to the Archival and Dedicated spectroscopic datasets, described in the text, and parts per million units are used for all variables.}
\end{figure*}

Figure \ref{fig3} and the bottom two rows of Table \ref{table2} show the analogous results to those of the previous section for the series truncated at cubic order. The first point to note is that we confirm both that the constraint on $a_1$ is unchanged, (with the posterior still recovering the prior) and the degeneracy between $a_2$ and $a_3$. For the model-independent cases the constraints on $a_2$ are not significantly changed, although its posterior likelihood for $a_2$ becomes quite non-Gaussian, while $a_3$ is unconstrained. For the full dataset the best-fit value for $a_2$ is slightly shifted towards negative values, which offsets a mild preference for a positive $a_3$. In any case the values of the reduced chi-square are still around $1.0$ for the archival dataset and around $0.8$ for the dedicated dataset, whether or not the high-redshift data  is used. In other words, they are not significantly changed with respect to those in the previous section, showing that statistically the additional parameter is not required.

In the results reported herein, we have assumed uniform priors for $a_2$ and $a_3$, in the range $[-25,+25]$ ppm. We should also note that the results in this section do have some dependence on the choice of priors for these two parameters. This is not surprising, being a well-known limitation of the cosmographic approach. In our case the effect seems to be milder than the one for the usual cosmography \cite{Dunsby}, which if true should be a consequence of the fact that $a_1$ is directly measured by atomic clock tests and therefore immune to this effect. While a full quantification of this dependence requires a more extensive treatment (including simulated data for cases with and without variations of $\alpha$), our results so far suggest that biases are minimized by choosing identical priors for $a_2$ and $a_3$.

\section{Conclusions}\label{concl}

We have introduced a cosmographic type analysis for local and astrophysical tests of the stability of the fine-structure constant, $\alpha$. This is an interesting context for a cosmographic approach for at least two reasons: the first term in the series is directly measured by laboratory experiments with atomic clocks, and astrophysical data exists in a wide range of cosmological redshifts all the way to the BBN epoch. Although not all of these measurements are model-independent, most of them have a typical sensitivity at the ppm level (with the atomic clock measurements being more stringent that this, the CMB one much less so), and can constrain the cosmographic series parameters to that same level of sensitivity.  

The results of our analysis are summarized in Table \ref{table2}. We note that we have not included the so calleg geophysical constraints from the Oklo natural nuclear reactor and meteorites in the low-redshift data since they are model-dependent: they are only bounds on alpha if one assumes that all other couplings are unchanged. Nevertheless, we have quantified the impact of adding the Oklo bound \cite{OKLO} to the analysis reportend in the first line of Table \ref{table2}. For the Archival dataset the constraint on $a_2$ would be unchanged (to the reported number of significant digits), while for the Dedicated one it would change from $-1.1\pm3.6$ to $-0.8\pm3.3$; the reason for this small impact is that although the Oklo bound is nominally quite stringent, it is also a very low redshift one.

Our results show that current data provides strong constraints on the first two terms in the series, while not warranting the addition of additional terms in the series. This is encouraging, because with such additional terms the $\alpha$ cosmographic series would become vulnerable to the same dependence of the choice of priors that affects the usual cosmographic analysis. In the case of $\alpha$, since we expect any putative variation to be slow and, for most physically motivated models, limited to the matter era, one might expect that a series truncated at quadratic order would suffice, with the $a_1$ parameter measured by local experiments and $a_2$ encoding model-specific information. These expectations need to be probed by a subsequent analysis, relying on simulated data for various fiducial particle cosmology models where $\alpha$ is redshift dependent. On the observational side, the arrival of next-generation spectrographs like ESPRESSO and ELT-HIRES will also push the sensitivity of astrophysical measurements beyond the parts per million level.

\section*{Acknowledgements}

This work was financed by FEDER---Fundo Europeu de Desenvolvimento Regional funds through the COMPETE 2020---Operational Programme for Competitiveness and Internationalisation (POCI), and by Portuguese funds through FCT - Funda\c c\~ao para a Ci\^encia e a Tecnologia in the framework of the project POCI-01-0145-FEDER-028987 and PTDC/FIS-AST/28987/2017. 

\bibliographystyle{model1-num-names}
\bibliography{alpha}
\end{document}